\documentstyle[12pt]{article}
\begin{document}
\thispagestyle{empty}
$\;$
\begin{center}
$$\;$$
{\large {\bf QUANTIZATION ON THE CONE AND \\ CYON--OSCILLATOR DUALITY}}
\end{center}
\begin{center}
{\bf A.N. Sissakian,
V.M. Ter--Antonyan}\footnote{E-mail: terant@thsun1.jinr.dubna.su} \\[3mm]

{\small N.N.~Bogoliubov Laboratory of Theoretical Physics,} \\
{\small Joint Institute for Nuclear Research,} \\
{\small Head Post Office, P.O.~Box 79, Moscow 101000, Russia}
\end{center}

\begin{abstract}
It is shown that the three--dimensional isotropic oscillator with coordinates
belonging to the two--dimensional half-up cone is dual to the
cyon , i.e. the planar particle--vortex bound system provided by
fractional statistics.
\end{abstract}

Our interest in the paper will be to consider
the three-- dimensional isotropic oscillator with coordinates belonging to
a two-- dimensional cone without the tip. We show that this system is dual to
the cyon [1], i.e. a planar particle--vortex bound system
with the Coulomb attractive interaction. This cyon is a simple prototype for
the objects with fractional statistics [2], known also as anyons [3]. Anyons
play an important role in the field theory [4], in the fractional quantum Hall
effect [5], and in high--$T_c$ superconductivity [6].

  Let $u_{\mu}$  and $(u,\theta,\varphi)$ be Cartesian and spherical
coordinates in ${\rm I\!R}^3$. Define points $u_{\mu}$ as belonging
to the half-up cone $C^{+}_2$, if $\theta = {\pi}/{6}$ and $u\neq 0$.
Consider the Schr\"odinger equation

\begin{eqnarray}
\frac{\partial^2\Psi}{\partial u_{\mu}^2}
 +\frac{2M}{\hbar^2}\left(E-\frac{M\omega^2u^2}{2}\right)\Psi=0,
\,\,\,\,\,\,\,\,{u_{\mu}\in C_2^+}
\end{eqnarray}
It is easy to verify that
\begin{eqnarray}
\left(\frac{\partial^2}{\partial u_{\mu}^2}\right)_{u_{\mu}\in C_2^+}
=\frac{4}{r}\left(\frac{\partial^2}{\partial r^2}+\frac{1}{r}
\frac{\partial}{\partial r}+\frac{1}{r^2}
\frac{\partial^2}{\partial\varphi^2}\right)
\end{eqnarray}
where $r=u^2$.

Combining (1) and (2), we obtain the Schr\"odinger equation in
${\rm I\!R}^2$
\begin{eqnarray}
\frac{\partial^2\Psi}{\partial r^2} + \frac{1}{r}\frac{\partial\Psi}
{\partial r} +\frac{1}{r^2}\frac{\partial^2\Psi}{\partial{\varphi}}
+ \frac{2M}{\hbar^2}\left(\frac{M\omega^2}{8}+\frac{E}{4r}\right)\Psi=0
\end{eqnarray}
The tip of the cone has been removed $(u > 0)$ and hence it is not possible
to deform one into the other two loops with different winding numbers.
As a consequence, $\Psi$ satisfies the twisted boundary condition [7]
\begin{eqnarray}
\Psi(r,\varphi+2\pi)=e^{i\pi\nu}\Psi(r,\varphi)
\end{eqnarray}
As a statistical parameter $\nu$  can be  arbitrary ($\nu\in [0,1]$),
our system is an anyon.In particular, if $\nu=0$ or $\nu=1$,  the
wavefunction picks up a plus or a minus sign, and the system acquires
bosonic or fermionic statistics respectively.

Let us introduce the new wavefunction $\psi$ instead of the previous one
\begin{eqnarray}
\Psi(r,\varphi)=\psi(r,\varphi)e^{i{\nu}{\varphi}/2}
\end{eqnarray}
Equation (3) transforms into
\begin{eqnarray}
\frac{\partial^2\psi}{\partial r^2} +
\frac{1}{r}\frac{\partial\psi}{\partial r} +
\frac{1}{r^2}\left(\frac{\partial}{\partial \varphi} +
i\frac{\nu}{2}\right)^2 \psi +
\frac{2M}{\hbar^2}\left(\frac{M\omega^2}{8}+\frac{E}{4r}\right)\psi=0
\end{eqnarray}
with the periodic boundary condition
\begin{eqnarray}
\psi(r,\varphi+2\pi)=\psi(r,\varphi)
\end{eqnarray}
We can see from (3) and (6) that the canonical angular momentum operator
$\hat J_c$=$-i{\partial}/{\partial\varphi}$ is replaced by the kinetic
angular momentum operator $\hat J=\hat J_c+s$. Here  $s={\nu}/2$ has the
meaning of the spin of our system. Thus, the spin $s$ and the statistics
$\nu$ appear to be related in the conventional way (${\nu}=2s$).

To clarify the physics associated with equation (6), let us introduce
the Cartesian coordinates
\begin{eqnarray}
x_1=r\cos{\varphi} \,\,\,\,\,\,\,\,\,\, x_2=r\sin{\varphi}
\end{eqnarray}
It is easy to verify by direct computation that (6) may be
rewritten as a Pauli equation
\begin{eqnarray}
\frac{1}{2M}\left(\hat p +\frac{e}{c}\vec A\right)^2\psi
-\frac{\alpha}{r}\psi=\epsilon\psi
\end{eqnarray}
with the vector potential
\begin{eqnarray}
\vec A(\vec x)=\frac{\Phi}{2{\pi}}\left(-\frac{x_2}{x_1^2 + x_2^2}\hat x_1
+ \frac{x_1}{x_1^2 + x_2^2}\hat x_2\right)
\end{eqnarray}
Here $\hat x_1$ and $\hat x_2$ are unit vectors along the $x_1$- and the
$x_2$-axes, ${\Phi}$=${\hbar}c{\nu}/2|e|$ and
\begin{eqnarray}
\epsilon=\frac{M\omega^2}{8},\,\,\,\,\,\,\,\,\,\,\,\,\,\,
\alpha=-eq=\frac{E}{4}
\end{eqnarray}
Thus, our initial system is equivalent to the cyon,i.e. the system
consisting of a point charge particle moving on the plane in the
external static electric and magnetic fields of a vortex localized at
the origin: $B$=${\Phi}\delta^{(2)}({\vec x})$. This system continuously
interpolates the bosonic $(s=0)$ and the fermionic $(s=1/2)$ systems
which have been obtained from the circular oscillator by the quantum
Bohlin transformation [8].

Let us introduce the separation ansatz
\begin{eqnarray}
{\psi}(r,\varphi)=R(r)e^{im{\varphi}}/{\sqrt{2{\pi}}}
\end{eqnarray}
Here $m$ is an integer because of the
boundary condition (6). This is thrue despite of the fact that the
algebra  of the two-dimensional rotations is abelian and in principle
an arbitrary constant could be added to the angular eigenvalues [9]
By substituting the wavefunction (12) into (6) we are led to the
radial equation
\begin{eqnarray}
\frac{d^2R}{dr^2} + \frac{1}{r}\frac{dR}{dr}
- \frac{(m+s)^2}{r^2}R + \frac{2\mu}{\hbar^2}\left(
\epsilon +\frac{\alpha}{r}\right)R = 0
\end{eqnarray}
and conclude that the radial function $R$ and the energy spectrum $\epsilon$
of the bound states of the cyon may be estimated from eigenfunctions and
eigenvalues of the two- dimensional Coulomb problem [10] by the substitution
$m\to{m+s}$.

Thus, we obtain
\begin{eqnarray}
R(r)=Ce^{-z/2}{z^J}F({-\sqrt{{\alpha}/{(-2{\epsilon}r_0})} +|m+s|+1/2}, {2J+1}, z)
\end{eqnarray}
where $z=2r/(-2{\epsilon})^{1/2}r_0$,  ${r_0}={{\hbar}^2}/M{\alpha}$,
$J=m+s$,  $s={\nu}/2$.

The corresponding energy is
\begin{eqnarray}
{\epsilon}^s_{n_{\rho}m} =-\frac{M{\alpha}^2}{2{\hbar}^2(n_r+
|m+s|+1/2)^2}
\end{eqnarray}
The value of the normalization constant
\begin{eqnarray}
C=\frac{2}{(-2{\epsilon})^{1/2}}\frac{1}{\Gamma(2J+1)}{\sqrt\frac{\Gamma
(2j+{n_r}+1)}{(2n_r+2J+1)({n_r})!}}
\end{eqnarray}
follows from the condition
\begin{eqnarray}
\int_{0}^\infty|R(r)|^2rdr=1
\end{eqnarray}
Equations (15) and (11) may be combined to yield
\begin{eqnarray}
E={\hbar}{\omega}{\left(2n_r+2|m+s|+1\right)}
\end{eqnarray}

We conclude that the energy levels for the isotropic oscillator on the cone
are identical with energy levels of the circular oscillator possessing,
apart from the usual angular momentum $m$, also the intrinsic (topological)
quantum number ${\nu}$ which gives rise to the spin $s={\nu}/2$ of the cyon.
The spin $s$ is localized near cyon and the kinetic angular momentum $J=m+s$
is located at the spatial infinity,as for the cyon without attractive
Coulomb interaction [11].

We are now in a position to give explanation for the meaning of the term
"duality" which we  have used in the title of our paper. Equations (1) and (9)
are connected with each other by the ansatz $(E,\omega) \to (\epsilon,\alpha)$
and by the transformation exchanging the coordinates $(u_1,u_2)$ and $(x_1,x_2)$
in the following way
\begin{eqnarray*}
x_1 =2uu_1,\,\,\,\,\,\,\,\,\,\,\,\,
x_2 =2uu_2
\end{eqnarray*}
Now,in equation (1) the coupling constant $\omega$ of the oscillator is fixed
and the energy $E$ is quantized. The situation with equation (9) is inverse:
here the cyon's coupling constant $\alpha$  (or $E$) is fixed and the energy
$\epsilon$ (or $\omega$) of the cyon is quantized. According to (11),these
conditions are inconsistent among themselves and, therefore, the isotropic
oscillator on the cone and the cyon are rather dual than identical to each
other [12].

\vspace{5mm}

We would like to thank Levon Mardoyan and
George Pogosyan for helpful discussion.

\end{document}